\documentclass[%
twocolumn,
superscriptaddress, 
floatfix,
 amsmath,amssymb, 
 ]{wlscirep}

\usepackage[utf8]{inputenc}
\usepackage[T1]{fontenc}

\usepackage{graphicx}
\usepackage{subfigure}
\usepackage{multirow}
\usepackage{comment}

\usepackage{cuted}
\usepackage{amsmath}
\usepackage{mathtools}
\usepackage{xcolor}
\usepackage{dcolumn}
\usepackage{bm}
\usepackage{lipsum}

\usepackage{xr}
\graphicspath{ {./} }

\title{Non-linear inhibitory responses enhance performance in collective decision-making}

\author[1,*]{David March-Pons}
\author[1]{Romualdo Pastor-Satorras}
\author[2,3]{M. Carmen Miguel}

\affil[1]{Departament de Física, Universitat Politècnica de Catalunya, Campus Nord B4, 08034 Barcelona, Spain}
\affil[2]{Departament de Física de la Matèria Condensada, Universitat de Barcelona, Martí i Franquès 1, 08028 Barcelona, Spain}
\affil[3]{Institute of Complex Systems (UBICS), Universitat de Barcelona, 08028 Barcelona, Spain}
\affil[*]{david.march@upc.edu}

\begin{abstract}
    The precise modulation of activity through inhibitory signals ensures that both insect colonies and neural circuits operate efficiently and adaptively, highlighting the fundamental importance of inhibition in biological systems. Modulatory signals are produced in various contexts and are known for subtly shifting the probability of receiver behaviors based on response thresholds. Here we propose a non-linear function to introduce inhibitory responsiveness in collective decision-making inspired by honeybee house-hunting. We show that, compared with usual linear functions, non-linear responses enhance final consensus and reduce deliberation time. This improvement comes at the cost of reduced accuracy in identifying the best option. Nonetheless, for value-based tasks, the benefits of faster consensus and enhanced decision-making might outweigh this drawback.
\end{abstract}

\begin{document}
\flushbottom
\maketitle

\section{Introduction}

The behavioral and signaling patterns of social animal groups~\cite{collective_animal_beh_book, animal_signals_book} have sparked extensive research into collective behavior and decision-making, primarily to understand the underlying mechanisms that drive these emergent properties~\cite{bose2017}. Inhibitory signals, in particular, play an essential role in social insects, fine-tuning collective decision-making and coordinating critical tasks such as house-hunting and foraging~\cite{nieh_stop_1993, pastor_piping_2005, lau_nieh_2010, nieh_negative_2010, seeley_stop_2012}. These inhibitory signals, often communicated through vibrations or tactile interactions, allow colonies to efficiently allocate resources and labor. For instance, in honeybees, stop signals can prevent the recruitment of additional foragers to poor or perilous food sources, thereby optimizing foraging efforts~\cite{nieh_stop_1993, pastor_piping_2005, lau_nieh_2010, nieh_negative_2010}. Similarly, during nest site selection, bees use stop signals to halt the promotion of less suitable sites, ensuring that the colony converges on the best available option~\cite{seeley_stop_2012}. By integrating these stop signals, social insects enhance their ability to make adaptive and robust decisions, ultimately supporting the survival and success of the colony. The fascinating social behavior of honeybees, including their intricate recruiting signaling patterns such as the waggle dance~\cite{Seeley2006amSci}, has inspired the design of decentralized decision-making algorithms~\cite{pais2013, reina_desing_pattern, reina2017, gray_multiagent_2018, Leonard2024}, and their application to robotic systems~\cite{valentini2017}. 

According to Nieh~\cite{nieh_stop_1993,nieh_negative_2010} and Pastor {\em et al.}~\cite{pastor_piping_2005}, during foraging tasks, honeybees' stop signals can alter the probability of waggle dancers ceasing their dance and leaving the nest, thereby reducing recruitment. However, dancers do not exhibit an immediate response to these signals. This feature is characteristic of modulatory signals, which are produced in various contexts and are known for subtly shifting the probability of receiver behaviors based on their response thresholds. Lau {\em et al.}~\cite{lau_nieh_2010} further suggested that, depending on receiver response thresholds, stop signals do not exert a strong colony-wide effect until signaling levels are sufficiently elevated. A similar mechanism has for long also been observed in brain neuronal activity~\cite{Wilson1972}, where balance between excitation and inhibition is critical for processes such as sensory perception, motor control or cognitive functions. Recent efforts have been made in establishing the similarities between individual decision-making in primate brains and collective decision-making in social insect colonies~\cite{passino_swarm_2008, marshall_optimal_2009, borofsky_hive_2020}. 

Field experiments on honeybee house-hunting~\cite{seeley_stop_2012} showed that stop signals are also pervasive in these scenarios. This study introduced the term \textit{cross-inhibition}, as it demonstrated that stop signals were predominantly exchanged between agents promoting competing options.
Cross-inhibition has proven essential for resolving deadlocks in decisions between very similar alternatives~\cite{seeley_stop_2012, pais2013, zakir_robot_2022, aust_hidden_2022, reina_cross_inhibition_2023}. 
However, as argued in Ref.~\cite{reina_cross_inhibition_2023}, cross-inhibition trades accuracy for stability. This means the system can confidently make a decision for any option, regardless of whether it is the highest quality one or not. In value-based tasks, this trade-off may not necessarily be detrimental, as the system prioritizes making a choice that yields a sufficiently high reward within a limited time, thus balancing the speed-value trade-off~\cite{passino_modeling_2006, pirrone2014, valentini2015_efficient9, reina_cross_inhibition_2023}. 
Furthermore, depending on the intensity of cross-inhibition, this mechanism may pause the decision-making process if the qualities of the available options are not deemed high enough, allowing the system to wait for a potentially better option to appear~\cite{pais2013, reina2017}. Such a system transitions from indecision to decision through pitchfork or saddle-node bifurcations~\cite{bizyaeva_nonlinear_2023, Leonard2024}, controlled by the model parameters.

In honeybee-inspired collective decision-making models, the cross-inhibition rate has usually been considered a linear function of the population sending the stop signals.
This choice represents the simplest modeling assumption, where the abandonment of one’s opinion is linearly proportional to the accumulation of stop signals received from peers with opposing options. However, similar to the foraging behavior of bees discussed earlier, Seeley et al. also suggested that nest-site scout waggle dances are likely terminated when stop-signal inhibition surpasses a certain threshold~\cite{seeley_stop_2012}.
Motivated by this experimental evidence, here we investigate the impact of non-linear inhibitory responsiveness \cite{talamali_improving_2019, bizyaeva_nonlinear_2023, Leonard2024} within honeybee-inspired decision-making models.
The response depends on the amount of stop signals received and diminishes or becomes negligible when stop signals are sparse, see Fig.~\ref{fig:ci_sample_bifus}(a).
This approach also aligns with the concept of complex social contagion models~\cite{centola2007complex,watts2002simple}, which posits that multiple exposures to a given opinion are required to trigger a shift in belief. Similarly, our model assumes that a minimum threshold of stop signals must be reached before initiating the cross-inhibition response.

Focusing on binary decision tasks, we demonstrate that our  approach enhances the consensus formation capabilities of decentralized systems compared to linear cross-inhibition models, particularly when dealing with options of similar qualities. The benefits are twofold: first, the final decision is achieved with virtually no bees committed to the less favored option; second, the time to reach a stationary state is significantly reduced. 

\section{Results and discussion}

\subsection{Model}
We use a simplified version of the model proposed by List et al. ~\cite{list2009}, referred to as the LES model, which is inspired by the house-hunting behavior of honeybees and serves as the basis of our decision-making framework. This agent-based model incorporates the typical discovery, abandonment, and recruitment transitions found in other collective decision-making models~\cite{pais2013, reina_desing_pattern, reina2017, gray_multiagent_2018}. 
Although the original model did not include cross-inhibition interactions, extending it to incorporate this mechanism is straightforward.

In the LES model, a swarm of $N$ scout bees,  indexed by $i = 1,\ldots, N$, evaluates $k$ potential nest sites,  indexed by $\alpha = 1,\ldots, k$. Each site $\alpha$ is characterized by an intrinsic quality $q_{\alpha} \geq 0$ and a spontaneous discovery probability $\pi_{\alpha} \geq 0$. 
%
Bees can be in any of $k+1$ states: uncommitted or committed to one of the $k$ available sites. 
The transitions from uncommitted to committed state are governed by discovery and recruitment rates, which represent individual and social behavior, respectively, and are balanced by an interdependence parameter $\lambda$.
Likewise, the transitions from committed to uncommitted states are governed by abandonment and cross-inhibition rates, which reflect individual and socially motivated behaviors, respectively. 

The model's mean-field rate equations for the fractions of agents committed to each site, $f_\alpha(t)$, can be derived using the master equation formalism~\cite{galla2010}. Including cross-inhibition, these equations are:
\begin{eqnarray} \label{eq:model_time_evo_linci}
    \dot{f}_{\alpha}(t) &=&  f_{0}(t) \left[(1-\lambda)\pi_{\alpha} + \lambda  f_{\alpha}(t)\right] \\
    &-& r_{\alpha} f_{\alpha}(t) - \lambda' f_{\alpha}(t) \sum_{\beta \neq \alpha} \sigma(f_\beta), \ \ \alpha=1,\ldots,k \nonumber
\end{eqnarray}
where $f_{0}(t) = 1 - \sum_{\alpha = 1}^{k} f_{\alpha}(t)$ is the fraction of uncommitted bees. The discovery rate, $(1-\lambda)\pi_\alpha$, refers to the rate at which uncommitted bees discover and commit to site $\alpha$, and the recruitment rate $\lambda f_\alpha$ represents the rate at which uncommitted bees are recruited by peers already committed to option $\alpha$. The rate $r_{\alpha}$ at which bees stop advertising a site is inversely proportional to its quality, $r_{\alpha}=1/q_\alpha$. Finally, the cross-inhibition rate, $\lambda' f_\alpha \sigma(f_\beta)$ ($\beta \neq \alpha$), is the rate at which bees abandon their options after receiving stop signals from those advocating for competing options. Here, $\lambda'$ regulates the intensity of cross-inhibition interactions. For the purposes of this study, we will set $\lambda' = 1$.

The stationary points of the system can be determined by numerically solving the equations obtained by setting $\dot{f}_{\alpha}(t) = 0$~\cite{marchpons2024consensus}. 
Without cross-inhibition ($\lambda' = 0$), the system simplifies to the expressions derived in~\cite{galla2010}. This particular system has been thoroughly analyzed in Refs.~\cite{marchpons2024_kilobots, marchpons2024consensus}.

\begin{figure}[t!]
    \centering
    \includegraphics[width=0.95\columnwidth]{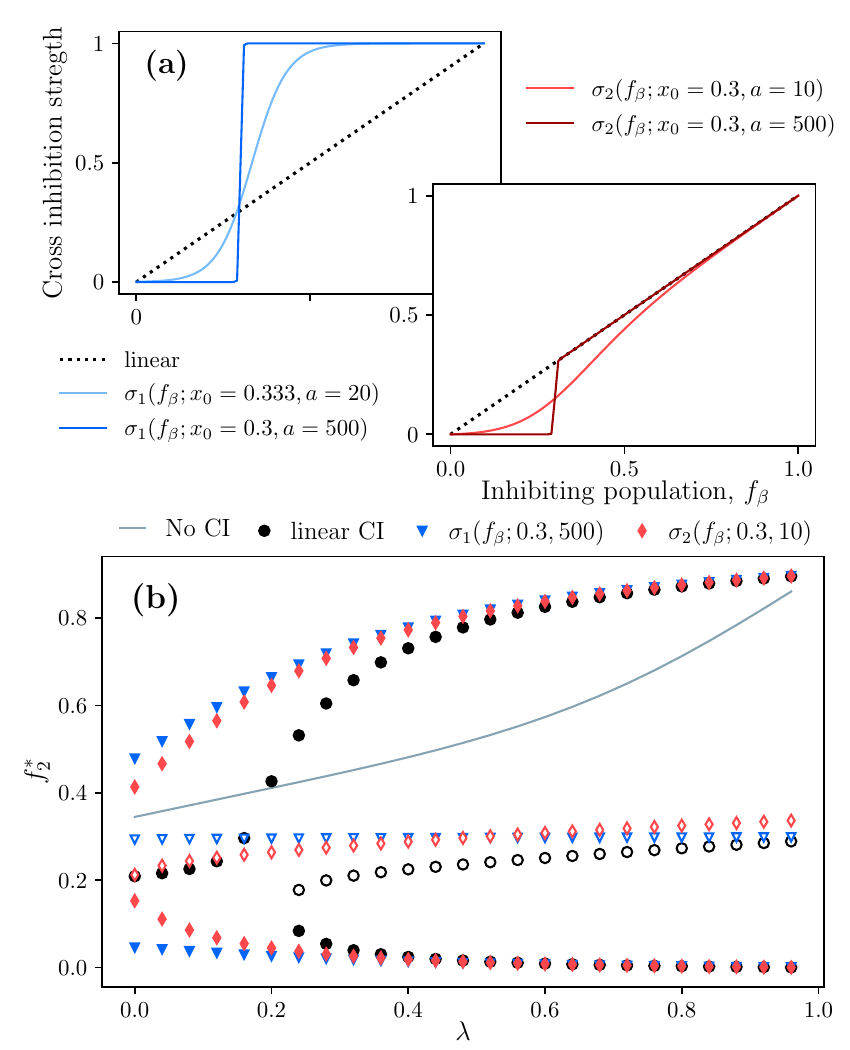}
    \caption{\textbf{Non-linear inhibitory responses and their effects on collective decision-making.} \textbf{(a)}: Strength of the cross-inhibition non-linear responses as a function of the population fraction that is sending the inhibitory signals, $f_\beta$. 
    Different lines represent responses that will be studied throughout the text, separated into two panels for clarity. On the left, a smooth sigmoid, $\sigma_1(f_\beta; x_0 = 0.333, a = 20)$ (light blue), and a sharp sigmoid $\sigma_1(f_\beta; x_0 = 0.3, a = 500)$ (dark blue), are depicted. On the right a smooth linearly bounded sigmoid, $\sigma_2(f_\beta; x_0 = 0.3, a = 10)$ (light red), and a sharp linearly bounded sigmoid $\sigma_2(f_\beta; x_0 = 0.3, a = 500)$ (dark red), are depicted. The parameter $x_0$ controls the ascent of the sigmoid and the parameter $a$ controls the smoothness of the ascent (see Eq.~\eqref{eq:sigmas} for more details). The black dotted line indicates a linear cross-inhibition response.
    \textbf{(b)}: Bifurcation diagrams on increasing interdependence $\lambda$ for linear cross-inhibition (black circles), a sharp sigmoid cross-inhibition function $\sigma_1(f_\beta;x_0 = 0.3,a=500)$ (blue squares), and a smooth bounded sigmoid function $\sigma_2(f_\beta;x_0=0.3,a=10)$}  (red triangles). The case $\lambda'=0$ is included as a continuous line for comparison. Other model parameters used are $\pi_1 = \pi_2 = 0.1$, $q_1 = 9$, $q_2 = 10$ and $\lambda'=1$.
    \label{fig:ci_sample_bifus}
\end{figure}

\subsection{Non-linear cross-inhibition response}
The function $\sigma (f_\beta)$ in Eq.~\eqref{eq:model_time_evo_linci} determines the actual strength of the cross-inhibition based on the fraction of bees $f_\beta$ sending stop signals. Traditionally, cross-inhibition, similarly to recruitment interactions, has been modeled as a proportional response to the fraction of adversary population, i.e. $\sigma(f_\beta) = f_\beta$.  Here we consider a non-linear cross-inhibition response. Specifically, we propose two sigmoid-like test functions, where the cross-inhibition strength remains weak for small values of the inhibiting population:
\begin{equation}
\begin{split}
    \sigma_1(f_\beta; x_0, a) = \frac{1}{1+e^{-a(f_\beta-x_0)}} \; \Theta(f_\beta),\\
    \sigma_2(f_\beta; x_0, a) = \frac{f_\beta}{1+e^{-a(f_\beta-x_0)}}.
\end{split}
\label{eq:sigmas}
\end{equation}
The parameter $a$ controls the steepness of these functions, and $x_0$ is a threshold controlling the sigmoid's ascent position. In $\sigma_1$, the Heaviside step function ($\Theta(x) = 1$ if $x \geq 0$, else $\Theta(x) = 0$) ensures that the cross-inhibition response is turned off when there is no inhibiting population ($f_\beta=0$).
Some instances of these functions, tested later on the nest-site selection dynamics, are depicted in Fig.~\ref{fig:ci_sample_bifus}(a). The function $\sigma_1$ captures the scenario where the cross-inhibition strength increases sharply, similar to a step function, once the threshold population $x_0$ is approached. On the other hand, $\sigma_2$ assumes that the cross-inhibition strength grows sub-linearly below this threshold and transitions to a limiting linear regime above it.
The threshold $x_0$ represents the population fraction at which cross-inhibition begins to have a significant effect. In this study, we have fixed $x_0$ to be approximately one third of the total population. This choice ensures that a sufficiently large committed population, formed through quality-sensitive communication, is established before cross-inhibition takes effect. A lower $x_0$ would favor the option that gains an early advantage due to random fluctuations, while a higher $x_0$ would delay cross-inhibition until interdependence is strong and a leading option has already emerged.
The sigmoidal functions shown in Fig.~\ref{fig:ci_sample_bifus}(a) have slightly different $x_0$ values to ensure that both non-linear functions surpass the strength of the linear model at the same population fraction. This adjustment allows for a fair comparison of their impact on the decision-making process.

\subsection{Fixed point analysis}
In the following, we will focus on the simplest case of a binary decision between two sites that differ in quality ($q_1 < q_2$).
The system's dynamics display a different number of stable points for different values of the model's parameter. Due to the analytical complexity of the model's equations, we resort to numerical methods to obtain the different fixed points (see Methods).

Increasing the strength of the social interactions leads to an (unfolded) pitchfork bifurcation~\cite{bifurcationTh} between one stable fixed point and two (asymmetric) stable fixed points separated by a saddle node. 
This behavior is shown in Fig.~\ref{fig:ci_sample_bifus}(b) for linear cross-inhibition (black-circle curve), and has been previously observed in similar models~\cite{pais2013, reina_desing_pattern, reina2017}. 
When switching to non linear cross-inhibition, a bifurcation still occurs, but its position depends on the specific non-linear cross-inhibition function chosen. Two examples of this are also shown in Fig.~\ref{fig:ci_sample_bifus}(b). The curves with blue squares and red triangles represent results for a sharp sigmoid function $\sigma_1(f_\beta;0.3,500)$ and a smooth linearly-bounded sigmoid function $\sigma_2(f_\beta;0.3,10)$, respectively.
Results obtained for other non-linear functions displayed in Fig~\ref{fig:ci_sample_bifus}(a) are shown in 
Supplementary Figure SF1~\cite{suppmat}.
Reducing the strength of cross-inhibition, or slightly varying the threshold parameter $x_0$, produces qualitatively similar results, though the positions of the bifurcations are shifted. For bifurcation plots at $\lambda'=0.5$, 
see Supplementary Figure SF2~\cite{suppmat}.

\subsection{Performance measure}
To assess the model's performance with non-linear cross-inhibition interactions, we numerically evaluate the stationary fixed point values,  focusing on the occupation fraction for the best-quality site, $f_2^*$. This quantity represents the decision accuracy of the system. However, as previously discussed, decision accuracy alone is not the only relevant variable a system seeks to maximize, especially in value-based decisions~\cite{pirrone_magnitude-sensitivity_2022, reina_cross_inhibition_2023, pirrone2014}. 

In scenarios where the available sites are similar in quality, it may be preferable to make a quick decision rather than spending a large amount of time to choose a slightly better site. Therefore, in addition to accuracy, we use agent-based stochastic simulations to measure two additional performance metrics: (i) the probability $P(f_2^*)$ of reaching the best option;
and (ii) the time $t_{ss}$ required to settle into this stationary state. See Methods for simulation details. These complementary quantities provide a comprehensive evaluation of the system's decision accuracy and speed performance.

\begin{figure}[t!]
    \centering
    \includegraphics[width=0.95\columnwidth]{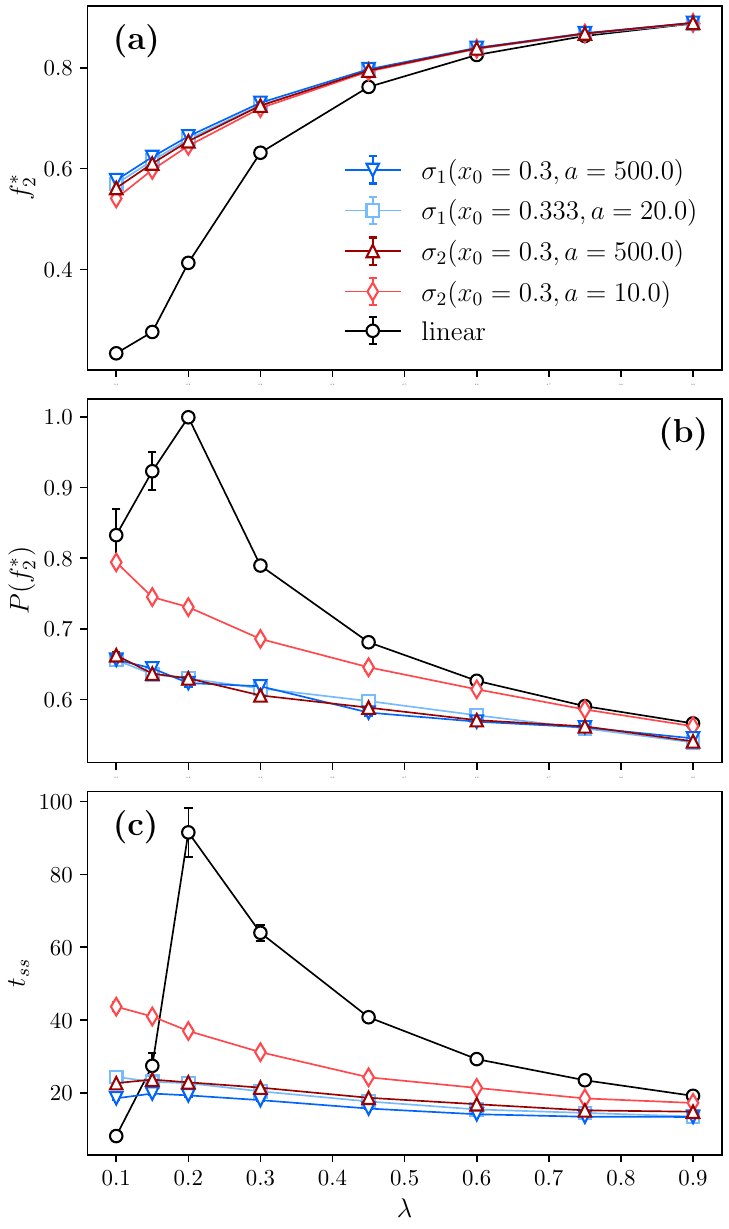}
    \caption{\textbf{Comparison between linear and non-linear cross-inhibition responses in a binary-decision problem.}
    \textbf{(a)}: Occupation fraction for the best-quality site, $f_2^*$. \textbf{(b)}: Probability of reaching the best option, $P(f_2^*)$. \textbf{(c)}: Time to settle into the stationary state, $t_{ss}$. Other model parameters are $\pi_1 = \pi_2 = 0.1$, $q_1 = 9$, $q_2 = 10$, and $\lambda'=1$, for a system of size $N=1000$. Error bars indicate the standard error of the mean.}
    \label{fig:costFuncData}
    \vspace*{-0.5truecm}
\end{figure}

Figure~\ref{fig:costFuncData} represents the behavior of these three quantities as a function of the interdependence  $\lambda$ for close values of the sites' qualities $q_1=9$ and $q_2=10$. Non-linear cross-inhibition results (see Eqs.~\eqref{eq:sigmas}) are shown together with the results of a linear cross-inhibition. In the non-linear case, we use the same parameters as in Fig.~\ref{fig:ci_sample_bifus}. 
We can observe that all non-linear cross-inhibition functions tested outperform the linear cross-inhibition in terms of pure consensus accuracy. However, the linear approach provides a higher probability of selecting the better option. These differences are particularly relevant for small to moderate values of the interdependence parameter, especially when the linear model has only one stable fixed point. 
In this regime, the system must balance independent discoveries and recruitment to build consensus for either option. Not triggering cross-inhibition unless an option has gained some representation allows the system to build a stronger consensus, albeit at the risk of less reliably choosing the better option. 
Nonetheless, this comes with the benefit of making a decision in a much shorter time, as shown in  Fig.~\ref{fig:costFuncData}(c). This can be a significant advantage when choosing between similarly valued options. As reported in~\cite{talamali_improving_2019, talamali2021_less_more}, quicker consensus can be achieved by allowing the system to first build sub-populations of comparable sizes before triggering competition between them. In those works, this is achieved by time-varying social interaction rates, including recruitment and cross-inhibition. In contrast, we propose a time independent mechanism that weakens the perception of cross-inhibition signals unless they are received from a significant portion of the population. This approach allows both populations to grow without interference from stop signals, either by pooling environmental cues or peer opinions. Once the populations reach substantial sizes, cross-inhibition is triggered, and a faster decision is made.

Each type of sigmoid function is tested with both a sharp response (high $a=500$ value), where the cross-inhibition rapidly shifts from no effect to maximum or linear bound, and a smooth response (low $a \in [10,20]$ value), where the transition to the final bound is more gradual. Interestingly, our results for the best-quality site occupation fraction show a remarkable insensitivity to the specific details of the sigmoid functions (see Fig.~\ref{fig:costFuncData}(a)). Moreover, these results are significantly higher for low and moderate interdependence compared to those of the linear cross-inhibition model.
On the other hand, the probability of retrieving the best option is considerably reduced for sharper cross-inhibitory responsiveness, independently of the function selected, Fig.~\ref{fig:costFuncData}(b). This is due to the indiscriminate action of inhibition on the option that first reaches the activation threshold $x_0$, irrespective of its quality. While the smooth sigmoid also yields probabilities similar to the sharp functions, due to the over-representation of the inhibiting population when the threshold is trespassed, approaching smoothly the linear bound grants an intermediate result.
It is also worth mentioning the significant reduction in deliberation time achieved with a non-linear cross-inhibition response, which occurs almost independently of the specific choice of non-linear function.

\begin{figure}[t!]
    \centering
    \includegraphics[width=0.95\columnwidth]{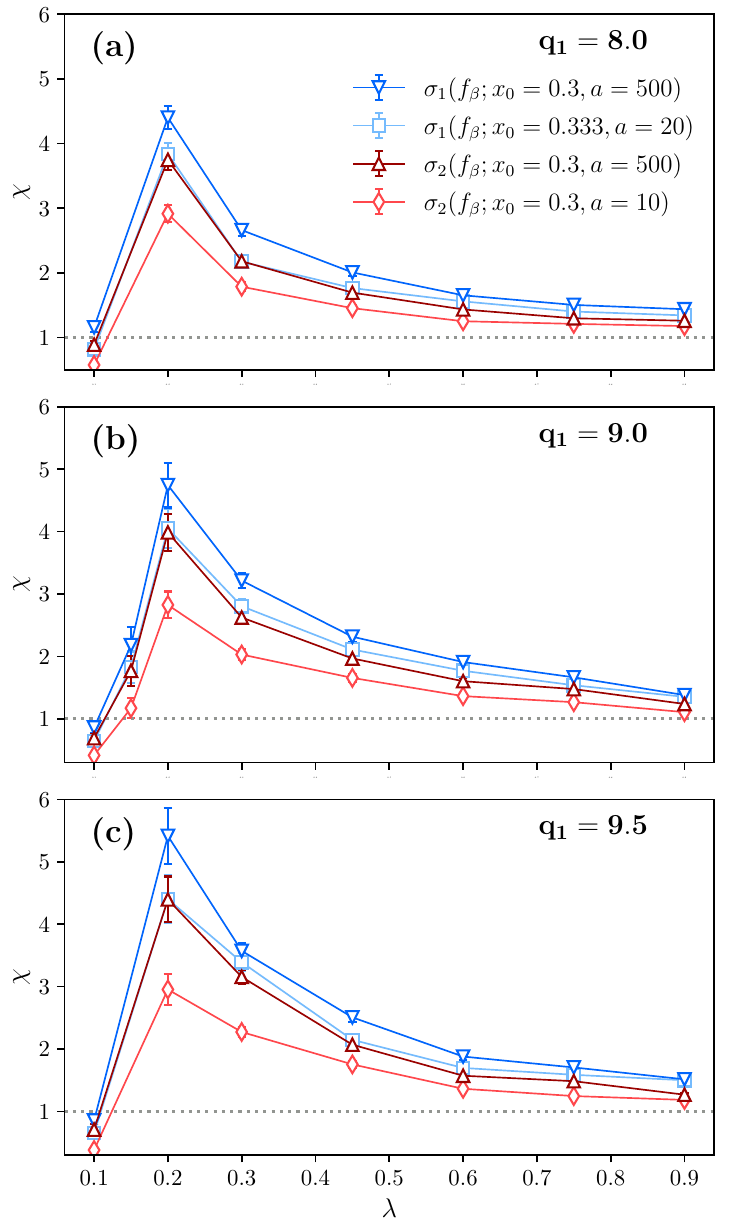}
    \caption{\textbf{Performance of non-linear cross-inhibition responses under varying interdependence.}
    Performance ratio $\chi$ of non-linear cross-inhibitory responses on increasing interdependence $\lambda$, in a binary-choice scenario. Three quality pairs are represented in \textbf{(a)}: $(q_1 = 8,q_2 = 10)$, \textbf{(b)}: $(q_1 = 9,q_2 = 10)$ and \textbf{(c)}: $(q_1 = 9.5,q_2 = 10)$. Other model parameters are $\pi_1 = \pi_2 = 0.1$, $\lambda'=1$ and system size $N = 1000$. Error bars indicate the propagated standard error of the mean.}
    \label{fig:chi_difqs_varl}
    \vspace*{-0.5truecm}
\end{figure}

In order to encapsulate the effect of these three measures in a single quantity that summarizes the performance of non-linear cross-inhibition, we define the objective performance, 
\begin{equation}
    \psi_{\sigma} = \frac{f_2^* \ P(f_2^*)}{t_{ss}},
\end{equation}
weighting the three quantities at stake. To assess how non-linear cross-inhibition compares to linear cross-inhibition, we also introduce the performance ratio $\chi = \psi_{\sigma}/\psi_{lin}$. 
Fig.~\ref{fig:chi_difqs_varl} depicts this performance ratio for three pairs of site qualities, ($q_1 = 8, 9, 9.5$ while $q_2 = 10$). Supplementary Figure SF3~\cite{suppmat} shows results for smaller values of $q_1$. In any case, we observe a performance ratio $\chi > 1$ for nearly all values of $\lambda$. Moreover, this ratio increases as the site qualities become closer, indicating a more significant performance improvement when using non-linear cross-inhibition.

The performance improvement peaks around $\lambda \sim 0.2$, corresponding to the point where the difference in decision times between the linear and nonlinear model is the greatest. As interdependence increases, the performance improvement diminishes because the three quantities become more similar across models. Nevertheless, non-linear responses still yield better overall performance.

The decrease in performance improvement with higher $\lambda$ is due to the combined effect of interdependence and cross-inhibition driving the losing population to very low fractions, while the winning population dominates (apart from a small uncommitted fraction). In this scenario, the cross-inhibition strength exerted by the winning population on its adversary becomes similar to that in the linear model, regardless of the specific non-linear response chosen. The advantage of the non-linear response is mainly due to the weaker effect of the losing population's cross-inhibition. Furthermore, as noted in~\cite{marchpons2024consensus}, when $\lambda \rightarrow 1$, the system can make a strong decision without cross-inhibition, although incorporating 
cross-inhibition significantly reduces decision time.

Comparing different non-linear cross-inhibition functions, we find that their performances are relatively close, with the smooth, linearly bounded sigmoid being the only one that underperforms. The effectiveness of a strong, sudden activation of cross-inhibition was previously reported by Talamali et al.~\cite{talamali_improving_2019}. However, while their study primarily noted an improvement in choosing accuracy without significantly affecting decision time, our approach demonstrates a comprehensive enhancement in both accuracy and decision time.

\begin{figure}[t!]
    \centering
    \includegraphics[width=.95\columnwidth]{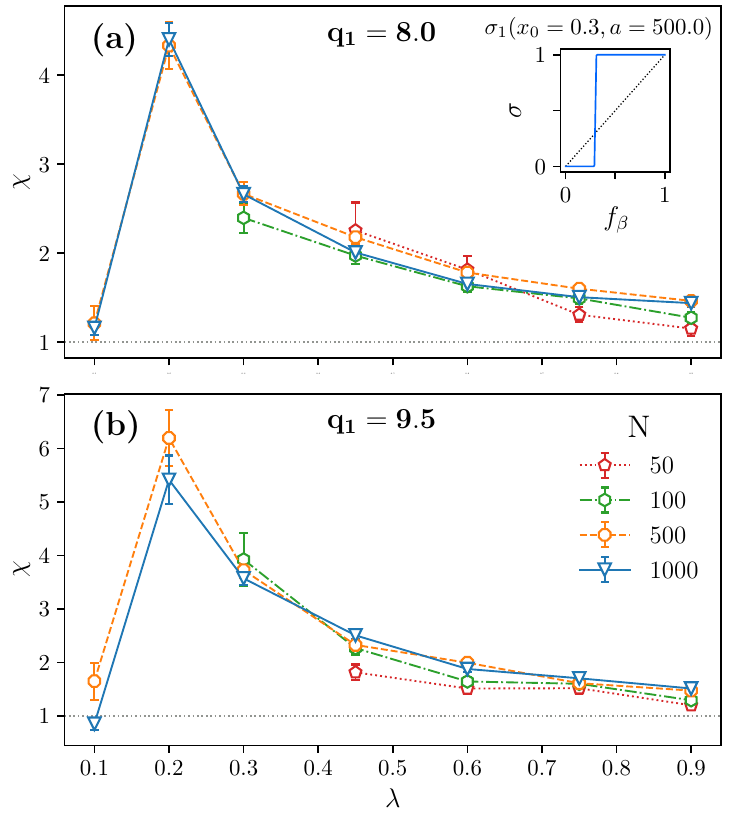}
    \caption{\textbf{Performance of a non-linear cross-inhibition response with varying system size.}
    Performance ratio $\chi$ of the sharp sigmoid non-linear cross-inhibition response $\sigma_1 (x_0 = 0.3, a = 500)$ on increasing interdependence $\lambda$ for different system sizes $N$.
    Other model parameters are $\pi_1 = \pi_2 = 0.1$, $\lambda'=1$, $q_2 = 10$, and $q_1 = 8$ (a) and $q_1 = 9.5$ (b), as indicated in each plot. Error bars indicate the propagated standard error of the mean. The inset in panel (a) shows the non-linear cross-inhibition function used, indicating the cross-inhibition strength $\sigma$ as a function of the inhibiting population $f_\beta$.}
    \label{fig:chi_var_N_sharp_sigmoid}
    \vspace*{-0.25truecm}
\end{figure}

One important question is the effect of system size on the relative performance of linear and non-linear cross-inhibition, as empirical studies suggest that colony size influences the effectiveness of cross-inhibition in real swarms (see, e.g., Ref.~\cite{Bell2021}). Smaller system sizes exhibit larger fluctuations, which can impact the ability of the swarm to consistently select the best option. To investigate this issue, we conducted numerical simulations for system sizes ranging from $N=50$ to $N=1000$. These simulations allow us to assess how fluctuations influence the probability of selecting the best option $P(f_2^*)$ and the time required to reach a stable consensus $t_{ss}$. In Supplementary Figure SF4~\cite{suppmat}, we show $P(f_2^*)$ and the time $t_{ss}$ for different system sizes. Finite-size fluctuations, which become more pronounced as the system size decreases, directly impact the probability of selecting the best option, as smaller systems are more susceptible to random fluctuations that can drive them toward an incorrect consensus state. In particular, these effects are more pronounced in the linear model. 
On the other hand, the effect on $t_{ss}$ is more nuanced. For small system sizes, the linear cross-inhibition model struggles to reach a true stationary state for all values of $\lambda$, particularly before or near the bifurcation point. Instead of stabilizing, fluctuations continuously drive the system between different consensus states indefinitely (see Supplementary Figure SF5~\cite{suppmat}). This explains why $t_{ss}$ cannot be properly identified for small $\lambda$ and $N= 50, 100$ in SF4 (right). As a consequence, for these values, $P(f_2^*)$ does not strictly measure the fraction of realizations in which the highest-quality option is chosen, and we discard them. In Figure~\ref{fig:chi_var_N_sharp_sigmoid} (R5), we plot the performance ratio as a function of $\lambda$ for different values of $N$. Other conditions are reported in Supplementary Figure SF6~\cite{suppmat}. Our results suggest that while finite-size effects play a role, particularly in smaller systems, non-linear cross-inhibition maintains its advantage by mitigating unwanted fluctuations more effectively than linear cross-inhibition.

\begin{figure}
    \includegraphics[width=.95\columnwidth]{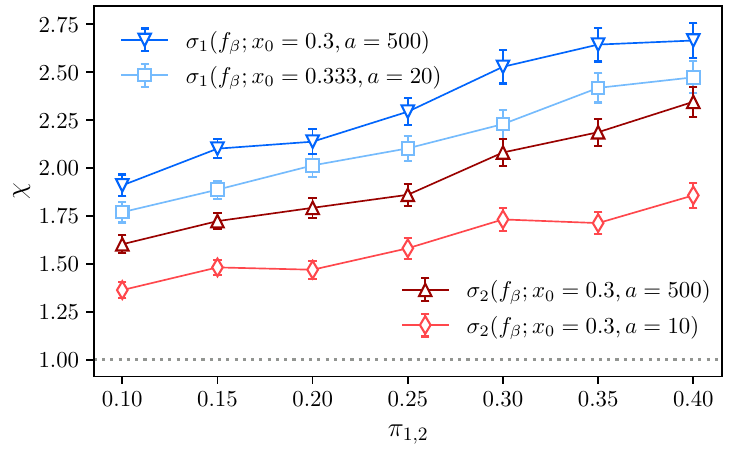}
    \caption{\textbf{Performance of non-linear cross-inhibition responses under varying options' discovery probabilities.} Performance ratio $\chi$ of non-linear cross-inhibition responses as a function of spontaneous discovery probabilities $\pi_1 = \pi_2  \equiv \pi_{1,2}$, in a binary-decision choice. Other parameters are $q_1 = 9$, $q_2 = 10$, $\lambda = 0.6$, $\lambda'=1$ and $N = 1000$. Error bars indicate the propagated standard error of the mean.}
    \label{fig:chi_varpi}
    \vspace*{-0.5truecm}
\end{figure}

The performance ratio allows us to asses the effect of different interaction patterns in various scenarios. So far, we have tested consensus dynamics by fixing the discovery probabilities and varying the swarm's interdependence. Increasing interdependence reduces the amount of individual exploration by prioritizing peers' options. This strategy has been shown to optimize consensus accuracy, even in the absence of cross-inhibition~\cite{list2009, marchpons2024consensus}, although it may extend decision time~\cite{marchpons2024consensus}. 
When the discovery probabilities increase (with fixed $\lambda$), the system more readily incorporates environmental information. 
This reduces decision time but leads to poorer final consensus, especially when options are of similar quality~\cite{marchpons2024consensus}.
The consensus decrease in this situation is caused by the fact that environmental information is socially unfiltered, i.e. it is incorporated at a rate that is independent of how much of the population is already advertising a given choice.
Thus, $\pi_\alpha$ can also be viewed as a noise parameter hampering overall accuracy. In such scenarios, cross-inhibition is crucial both to maintain high group cohesion (most of the population committed to the same option) and to avoid deadlocks, as reported in various case studies~\cite{pais2013,reina_desing_pattern,zakir_robot_2022,reina_cross_inhibition_2023}. 

Fig.~\ref{fig:chi_varpi} shows the performance ratio of non-linear cross-inhibition responses as discovery probabilities increase, for fixed values of quality and interdependence. The corresponding performance variables are plotted in Supplementary Figure SF7~\cite{suppmat}. As the noise in the system increases, non-linear cross-inhibition yields better performance. Weakening the stop signals from the losing population, whichever its quality, becomes essential in this context: although uncommitted agents continuously introduce "incorrect" information at a steady rate, the stop-signaling mechanism ensures that a dominant option can suppress this noise, thereby maintaining a high level of consensus. 
Interestingly, examining the individual quantities $f_2^*$ and $t_{ss}$ on increasing $\pi_1 = \pi_2 = \pi_{1,2}$,
we observe opposing trends for the linear and non-linear model. The linear response yields decreasing $f_2^*$ while increasing $t_{ss}$; in contrast, non-linear responses reverse this trend. Consequently, the model performance improves as $\pi_{1,2}$ increases. Comparing the different non-linear responses, we observe that the performance is consistently higher for the standard sigmoid functions than for the linearly bounded sigmoids. In each case, the sharp response also grants better performance.

Although this study focuses on a binary decision process, the model can be extended to scenarios involving more than two alternatives~\cite{reina2017}. In such cases, the advantages of non-linear models become even more pronounced, as demonstrated in Supplementary Figure SF8~\cite{suppmat} for a five-option scenario. When cross-inhibition is linear, the system settles into a stationary deadlock, with no single option dominating. In contrast, with non-linear cross-inhibition, the system initially appears stuck in indecision. However, due to stochastic fluctuations, one option eventually gains an advantage, leading to an asymmetric distribution of commitment. This outcome stems from the reduced inhibitory effect of smaller populations: in the linear model, all options exert mutual inhibition, preventing a clear decision. By contrast, non-linear cross-inhibition allows one option to exert a disproportionately strong inhibitory signal once it gains enough quorum, ultimately breaking the deadlock and establishing consensus.

Finally, it is worth noting that non-linear inhibitory responses can be broadly applied to various modeling scenarios. For instance, Reina et al.~\cite{reina_cross_inhibition_2023} propose a model where quality explicitly modulates the strength of recruitment and cross-inhibition interactions, while options' discovery and abandonment are considered noise, a quality-independent parameter. Modulating cross-inhibition interactions while varying this noise parameter produces results very similar to Fig.~\ref{fig:chi_varpi}.

\section{Conclusions}
Here we investigate non-linear cross-inhibition interactions in decentralized decision making models inspired by house-hunting honeybees. The primary design goal is to weaken an individual's response to stop signals when they are received from a small fraction of the population. We model this behavior using two non-linear functions, tested with different parameters (Fig.~\ref{fig:ci_sample_bifus}).
Focusing on binary decision tasks, we demonstrate that non-linear cross-inhibition results in higher consensus (the fraction of the population committed to the chosen option) and quicker decisions. These two benefits come at the cost of reducing accuracy in reliably choosing the best quality option. Nonetheless, in decisions made among options with close qualities, a stronger and quicker decision for a "good enough" option may be more beneficial than a weaker consensus or a slower decision process that yields the absolute best option~\cite{pirrone2014, pirrone_magnitude-sensitivity_2022, reina_cross_inhibition_2023}.
Moreover, while we have focused on scenarios where a single best option exists, both linear and non-linear models can also break symmetry between equally valued alternatives. In such cases, our key findings—namely, the superior performance of non-linear models in terms of consensus formation and decision time—remain valid.
Our results thereby open promising avenues for future research in decentralized collective decision-making and practical applications in swarm robotics. In this context, the predictions of our mean-field analysis could be tested through experiments with robot swarms, which can be easily programmed to respond non-linearly to stop signals from their nearest neighbors.

\section{Methods}
\subsection{Numerical analysis of the LES model with cross-inhibition}

\noindent\textbf{Linear Cross-inhibition.} When the cross-inhibition response is linear, the set of equations Eq.~\eqref{eq:model_time_evo_linci} can be combined to derive an equation for the fixed points of the uncommitted population. In each equation, the cross-inhibition term can be reformulated as $-\lambda f_\alpha \sum_{\beta \neq \alpha} f_\beta = -\lambda' f_\alpha (1 - f_0 - f_\alpha)$. By setting $\dot{f}_\alpha = 0$, an expression for $f_\alpha^*(f_0^*)$ can be obtained (Eq.~\eqref{eq:linci_falpha}). First, summing over all $\alpha$ yields a closed equation for the uncommitted population's fixed point, $f_0^*$ (Eq.~\eqref{eq:linci_f0}). This leads to the following set of equations:

\begin{strip}
    \rule{\dimexpr(0.5\textwidth-0.5\columnsep-0.4pt)}{0.4pt}%
    
    \begin{equation}
    1-f_0^* = -\frac{k(\lambda+\lambda')f_0^*}{2\lambda'}+\frac{k}{2} +
    \frac{1}{2\lambda'} \sum_{\alpha = 1}^k 
    \left[ r_\alpha  \pm \sqrt{((\lambda+\lambda')f_0^* -r_\alpha-\lambda')^2 - 4\lambda'(1-\lambda)f_0^*\pi_\alpha} \right]
    \label{eq:linci_f0}
    \end{equation}
    
    \begin{equation}
    f_\alpha^* = \frac{-((\lambda+\lambda')f_0^* - r_\alpha - \lambda') \pm \sqrt{((\lambda+\lambda')f_0^* - r_\alpha - \lambda')^2-4\lambda'f_0^*(1-\lambda)\pi_\alpha}}{2\lambda'} \qquad \quad \alpha = 1, ..., k
    \label{eq:linci_falpha}
    \end{equation}

    \par
    \hfill
    \rule[0.5\baselineskip]{\dimexpr(0.5\textwidth-0.5\columnsep-1pt)}{0.4pt}
\end{strip}

\noindent where $k$ is the number of sites. This expression provides as many equations as there are possible choices for the $\pm$ sign in Eqs.~\eqref{eq:linci_falpha}, which must be solved numerically. Not all sign combinations will yield a solution, but for those that do, we can determine the fixed points $f_0^*$ and subsequently compute the values of $f_\alpha^*$. The stability of these fixed points can by analyzed through a Linear Stability Analysis of Eqs.~\eqref{eq:model_time_evo_linci}, leading to the following expression for the elements of the Jacobian matrix:
\begin{eqnarray}
    J_{\alpha \alpha} &=& \lambda (f_0^*-f_\alpha^*) - \lambda'(1-f_0^*-f_\alpha^*)-(1-\lambda)\pi_\alpha - r_\alpha \nonumber \\
    J_{\alpha \beta} &=& -(1-\lambda)\pi_\alpha - f_\alpha^*(\lambda'+\lambda)  \nonumber \text{.}
\end{eqnarray}

\noindent\textbf{Non-linear Cross-inhibition. } 
When incorporating the non-linear cross-inhibition responses (Eq.~\eqref{eq:sigmas}),
it is not feasible to derive a closed-form equation for $f_0^*$, as is possible in the linear case. Instead, one must numerically solve the system of equations where $\dot{f}_{\alpha} = 0$. To find all fixed points, it is necessary to explore a sufficient number of points within the simplex $(f_\alpha; \; \alpha=1,...,k$) as initial guesses for the numerical solver. The stability of these fixed points can then be confirmed by numerically integrating the system's equations in the vicinity of the obtained solutions.

\subsection{Master equation and Stochastic Simulation Algorithm}
We employed Gillespie's algorithm~\cite{gillespie2013perspective} to estimate the stationary probability distributions, enabling us to analyze the likelihood of selecting the best site. The transition rates that define the master equation are inferred from the system's ODEs (Eqs.~\eqref{eq:model_time_evo_linci}). The transition rates used in the Gillespie algorithm are:
\begin{eqnarray}
   & & T_\alpha^{disc} = n_0(1-\lambda)\pi_\alpha, \text{\ discovery of option-$\alpha$,}\nonumber \\
   & & T_\alpha^{aban} = n_\alpha r_\alpha, \text{\ abandonment of option-$\alpha$,}\nonumber \\
   & & T_\alpha^{rec} = \frac{n_0 \lambda n_\alpha}{N}, \text{\ recruitment of option-$\alpha$,}\nonumber \\
  & &  T_\alpha^{c.i.} = n_\alpha \lambda' \sum_{\beta \neq \alpha} \sigma\left(\frac{n_\beta}{N}\right), \text{\ cross-inhibition of option-$\alpha$}. \nonumber 
 \end{eqnarray}
\noindent Here, $r_\alpha = q_\alpha^{-1}$ represents a simplified rate compared to the initial formulations of this model~\cite{list2009, galla2010}. The mathematical derivation that leads from the master equation to the ODE system (for the version of the model without cross-inhibition) can be found in detail in~\cite{galla2010}.

To estimate the stationary probability distributions, we run $10^4$ simulations of the stochastic simulation algorithm, setting a maximum time $t = 500$ to ensure a stationary state is reached. We then count how many realizations settle on each of the possible stable fixed points. However, when there is only one fixed point and the stationary values of the population fractions $f_1^*$, $f_2^*$ are very similar -- such as in the case of the linear cross-inhibition model for  $\lambda < 0.2$-- we use a different approach to obtain a more reliable estimate. In this case, we run $10^3$ longer simulations ($t = 10^4$) and collect $10^3$ evenly spaced data points from the stationary state. Using these values $(f_1^*, f_2^*)$, we compute the probability of each option winning.
For the stationary times, $t_{ss}$, we analyze $10^3$ of these trajectories to determine when the system reaches the stationary plateau. We bin the temporal evolution into intervals of approximately 1 unit of the dimensionless time variable and compare the absolute difference between the population values at each time and the stationary average (computed using the system state at sufficiently long times) for each population $\alpha = 0, ..., k$. When the difference satisfies $|f_\alpha(t) - \langle f_\alpha \rangle| < 0.1$, we consider the population to have entered the stationary state at time $t$. An estimate of $t_{ss}$ is then obtained by averaging the times at which this condition is met for each population (see Supplementary Figure SF9~\cite{suppmat}).

\section*{Data availability}
This study did not involve the analysis of new data sets.

\section*{Code availability}
The code with the stochastic simulation algorithm can be accessed at: \texttt{https://github.com/dmarchp/nonlinCI/tree/main}.

\vspace{0.5cm}

\section*{Acknowledgments}
We acknowledge financial support from projects PID2022-137505NB-C21 and PID2022-137505NB-C22 funded by MICIU/AEI/10.13039/501100011033, and by “ERDF: A way of making Europe”.

\section*{Author contributions}

D.M.-P., R.P.-S. and M.-C.M. designed the research. D.M.-P. developed the theoretical analysis and  numerical simulations. All authors analyzed the results and wrote the paper.

\section*{Competing interests}

The authors declare no competing interests.

\section*{Additional information}
Supplementary information for this paper is available at \texttt{link-to-be-inserted}.

\end{document}